\documentclass[11pt,a4paper]{article}
\usepackage{jinstpub}
\usepackage[export]{adjustbox}
\usepackage{makecell}

\DeclareMathOperator\erf{erf}

\title{On the depletion behaviour of low-temperature covalently bonded silicon sensor diodes}

\author[1]{J.~W\"uthrich%
\note{Corresponding author.}}
\author{and A.~Rubbia}

\affiliation{Institute for Particle Physics and Astrophysics, ETH Z\"urich,\\
Otto-Stern-Weg 5, Z\"urich, Switzerland}

\emailAdd{jwuethri@phys.ethz.ch}

\abstract{
Low temperature covalent direct wafer-wafer bonding allows for the fusion of multiple semiconductor wafers without any additional material at the bonding interface.
In the context of particle pixel detectors this might provide an alternative to bump-bonding for joining sensors to readout chips.
Previous investigations have shown that the amorphous layer formed at the interface during bonding is detrimental to charge propagation.
To investigate the influence of the bonding interface on signal collection we have fabricated custom test structures by bonding high-resistivity N to high-resistivity P-type silicon wafers thus forming P-N junctions.
Scanning transmission electron microscopy shows indeed the formation of ca. 3nm wide amorphous layer at the interface.
Using a scanning transient current technique (TCT) setup we were able to record generated signals.
Illuminating our sample with light of different wavelengths and from different sides, indicates that the P side of the bonded structures can be fully depleted, but not the N side.
This indicates a strongly asymmetric depletion behaviour which we attribute to the presence of the bonding interface.
}


\begin{document}

\maketitle
\flushbottom

\section{Introduction}
One of the challenges of building highly efficient X-ray imaging detectors, is the stochastic interaction of X-ray photons with matter dominantly via the photoelectric effect. This applies to the X-ray energy ranges commonly used in medical X-ray applications, ranging up to $100 \text{ keV}$.
Within this range, the probability of an X-ray photon being absorbed within a certain detector volume is closely related to the atomic number of the material used and its density.
Silicon (Si) is the material of choice for implementing particle / radiation pixel detectors, due to the availability and relative affordability of silicon semiconductor processing. But from the perspective of X-ray interaction silicon is not an ideal detector material \cite{seltzer_tables_1995}. This is for instance the case for X-ray photons with an energy $E_\gamma = 30 \text{ keV}$ which lies in the energy range often used for medical mammography examinations~\cite{jochelson_contrast-enhanced_2021}.
For medical applications, there is strong interest in having efficient pixel detectors, as this allows to obtain similar or higher quality images while at the same time exposing the patient to less radiation.
In order to have a near 100\% detection efficiency with a pure silicon detector, the interaction layer would need to be $2 \text{ cm}$ thick. Such thick detectors are impossible to manufacture with modern silicon semiconductor fabrication processes. The standard wafer thickness for a $200 \text{ mm}$ silicon process is ca. $700 ~ \mu\text{m}$.
This motivates the interest in different types of semiconductor interaction materials such as gallium-arsenide (GaAs) or cadmium-telluride (CdTe), also called high-Z materials. Highly efficient X-ray detectors can be built with reasonably thin layers of these alternative semiconductor materials.
But this only concerns the interacting part of an X-ray pixel detector. For economic reasons it is impossible to fabricate the integrated electronic circuits necessary for pixel detectors in GaAs or CdTe.
In order to build a highly efficient X-ray pixel detector it is extremely convenient to combine readout electronics implemented with standard silicon processes together with interaction layers made of high-Z materials.
It is possible to build such detectors, based on the hybrid manufacturing approach. Hybrid detectors are created by separately fabricating the interaction layer (called the sensor) and the readout electronics. The two parts are then joined together used bump bonding~\cite{delpierre_history_2014}. Due to the separate processing of the sensor part which usually consists of a simple implanted P-N junction, it is possible to create hybrid detectors, where the sensitive layer is made of a high-Z material~\cite{delpierre_history_2014}.
We investigate a novel approach for assembling high-Z particle pixel detectors based on a low-temperature covalent wafer-wafer bonding process.
Low temperature covalent wafer-wafer bonding is routinely utilized in the production of micro-electromechanical systems (MEMS) to encapsulate structures~\cite{esashi_direct_2021}, as well as in the production of high efficiency solar cells~\cite{predan_direct_2020}.
This bonding process enables the fusion of semiconductor materials without any additional material at the interface. The process allows to bond different materials together, such as bonding a GaAs wafer to a Si wafer. In the ideal case this would lead to abrupt hetero-junctions at the bonding interface. As the bonding process is carried out at or near room temperature, it is compatible with fully processed CMOS wafers. Materials with different thermal expansion coefficients can also be considered. This would allow to combine readout circuits fabricated using standard silicon processes with a wide range of different interaction materials, such as Ge, GaAs or CdTe.
The bonding is carried out at a wafer level, which enables mass processing compared to bump-bonding which usually occurs on a chip basis.

\begin{figure}
  \centering
  \includegraphics[width=\textwidth]{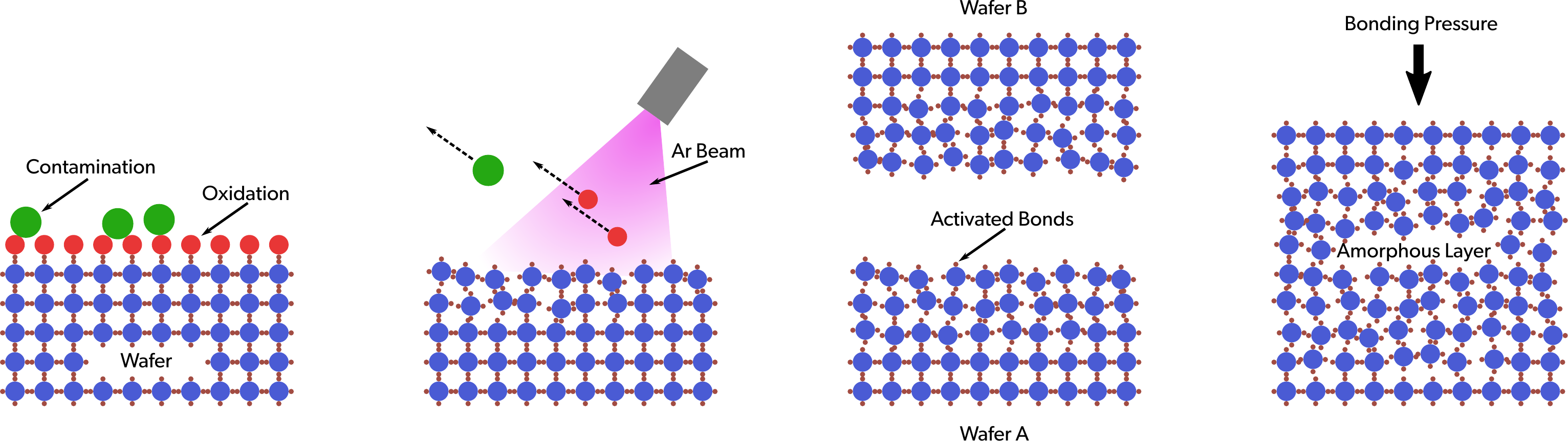}
  \caption{Conceptual view of the different steps of the surface activated covalent bonding method. Processing is carried out in high-vacuum ($< 5\cdot10^{-8} \text{ mbar}$). Contamination and oxide layers at the surface are removed using an argon beam, leaving behind activated dangling atom bonds. Bringing together two activated wafers leads to spontaneous bonding without the use of high temperatures.}
  \label{fig:covalent_bonding}
\end{figure}

The covalent bonding process is based on the principle of spontaneous covalent bonding between two wafers with dangling bonds at their surface~\cite{takagi_surface_1996}. The bonding process involves the following steps, which are all carried out under ultra-high vacuum ($p < 5 \cdot 10^{-8} \text{ mbar}$): both wafers are introduced into the vacuum and the surface of the first wafer is sputter-cleaned via either an argon beam or an argon plasma~\cite{esashi_direct_2021}\cite{neves_towards_2019}.\footnote{While argon is the most commonly used gas for surface activated bonding, the use of different noble gases such as neon has also been explored~\cite{esashi_direct_2021}.}
The sputter-cleaning removes any contaminants at the surface of the wafer as well as the thin native oxide layer formed on top of the silicon crystal. Due to the removal of the oxide layer, dangling (or activated) silicon bonds are generated at the surface. The second wafer is treated in the same way as the first one.
The activated surface of both wafers are then brought in contact. Given that the entire process is carried out in ultra-high vacuum, very little to no re-oxidation of the activated surfaces can occur. Therefore the activated bonds at the two surfaces spontaneously bond under moderate pressure. The resulting wafer-wafer bond is oxide-free and the mechanical bond strength approaches or equals the semiconductor bulk strength~\cite{esashi_direct_2021}.
High-temperature annealing is unnecessary for achieving high bond strengths, which ensures the compatibility of the process with fully processed CMOS wafers~\cite{neves_towards_2019}.
Figure~\ref{fig:covalent_bonding} shows a schematic view of the bonding process.
The surface activation via argon beam or argon plasma does not only remove the native oxide and contaminants, but it also attacks the surface of the semiconductor to be bonded. This leads to damages in the crystalline structure of the semiconductor and by extension to an amorphous layer at the location of the bonding interface. The thickness of this bonding interface is typically of the order of a few nano-meters, but amorphous layers as thin as $0.5 \text{ nm}$ have been reported in literature~\cite{neves_towards_2019}.
The result of this amorphous layer is the presence of bulk defects in the region around the bonding interface. These bulk defects influence the electrical properties of the bonded structure. When using covalent wafer bonding for steady-state applications (for example solar cells), the effects of the bulk defects can be mitigated by including high doping concentrations at the interface~\cite{predan_effects_2017}. This type of mitigation is not directly applicable to the fabrication of pixel detectors, since having highly-doped concentrations in the interaction layers will increase the necessary voltage for fully depleting the sensitive volume. Additionally, in the context of pixel detectors, the steady-state effects of the bulk defects are secondary. Our main interest lies with the transient effects occurring when a signal pulse crosses the bonding interface.

So in order to build efficient detectors in the future based on low-temperature direct wafer-wafer bonding, the effects of the defects at the bonding interface need to be studied in detail. Previous work in~\cite{bronuzzi_transient_2018} has investigated the usage of Si/Si wafer bonds for pixel detectors with Schottky diode based samples. The bonded wafers had the same doping type. Using transient current technique (TCT) measurements it was observed that the bonding interface blocks the passage of generated charge clouds. In~\cite{jung_sige/si_2018} p-Si/p-Si wafer bonded structures were fabricated, and the forward and backward bias currents were analysed. Again an influence of the bonding interface was observed reducing the conductivity of the bonded structure.
In both of these works, an alternative approach was proposed which might mitigate these observed problems. This alternative approach is to bond wafers of different types, so bonding a P-type to an N-type wafer, therefore creating a P-N junction located at the bonding interface. Under reverse bias, this structure would start depleting from the bonding interface, thus ensuring a high electric field at the bonding interface which is expected to facilitate charge transport across the interface.
This approach is studied in the present work. For simplicity reasons, as in the cited publications, this first investigative attempt is done with structures entirely composed of silicon. Once the \emph{simple} case of silicon bonded to silicon is understood, future investigations will analyse the creation of hetero-junctions, for example GaAs bonded to silicon.

\section{Pad diode test structures fabrication}
\begin{figure}
  \centering
  \includegraphics[width=0.66\textwidth]{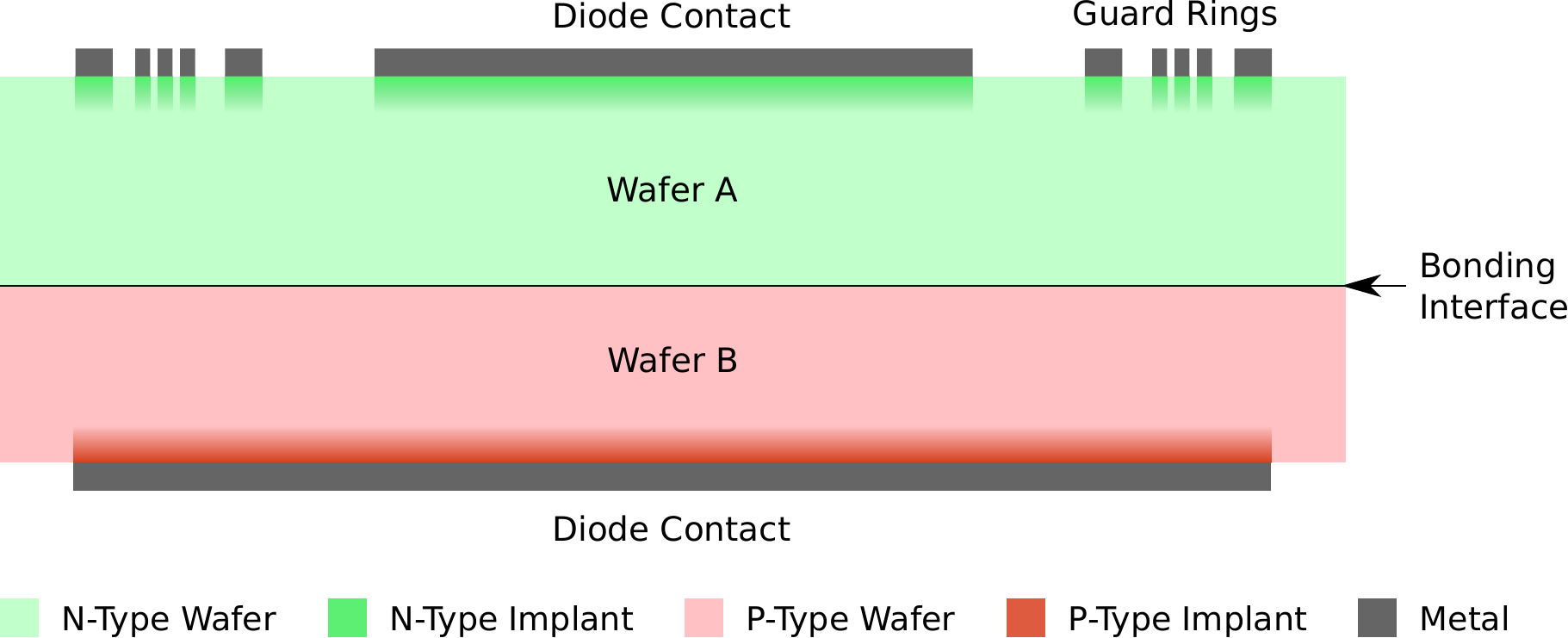}
  \includegraphics[width=0.32\textwidth]{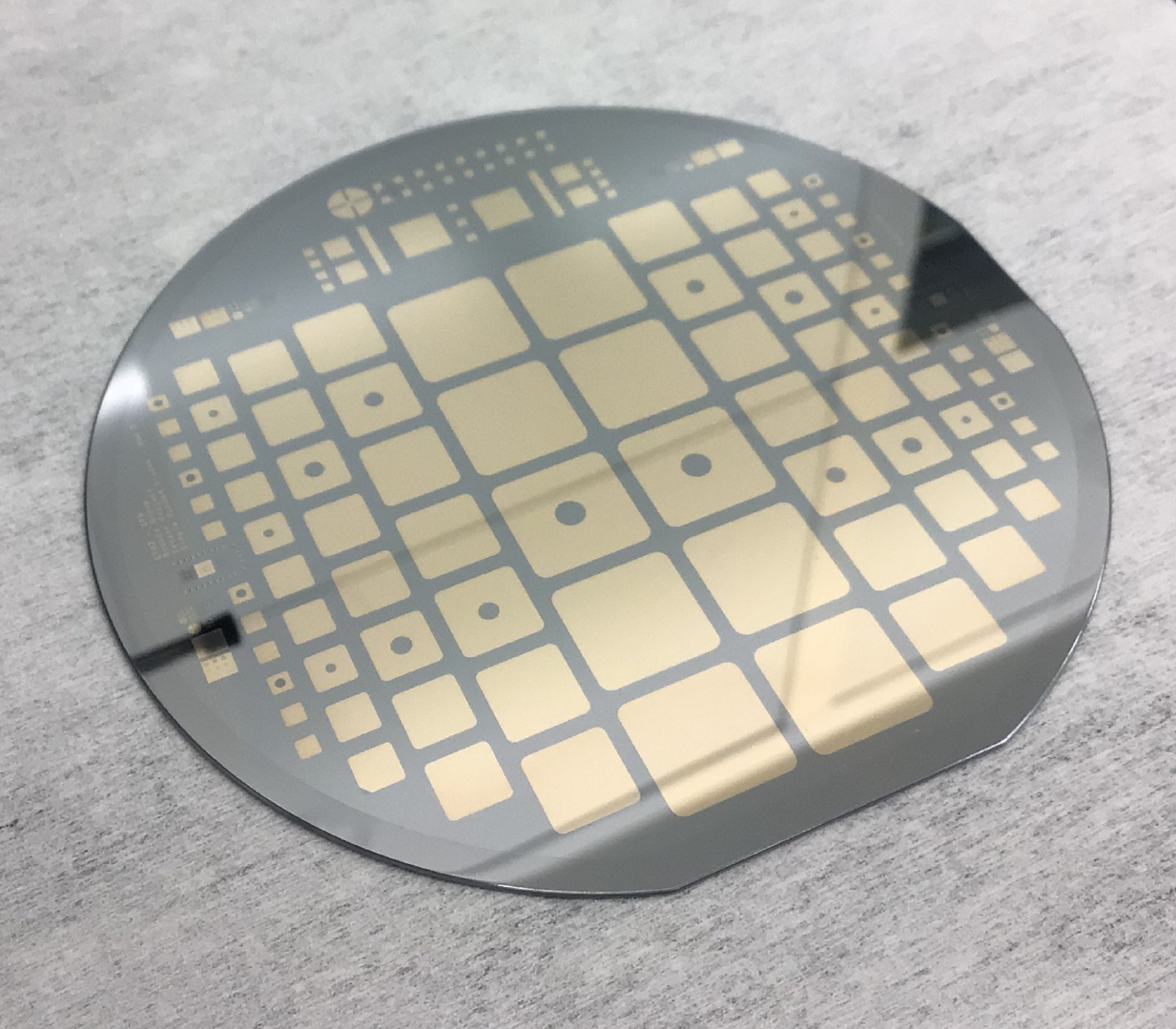}
  \caption{Left: Schematic view of the cross-section of the fabricated test structures. A high resistivity N-type silicon wafer is bonded to a high resistivity P-type wafer. Each wafer has high-dose implants at the surface in order to ensure good ohmic conduction between the silicon substrate and the metal contacts. Image not to scale. --- Right: P side of a fully processed wafer pair.}
  \label{fig:structure_section}
\end{figure}

To evaluate the influence of the bonding interface on charge transport and signal collection, we fabricated custom test structures. Each test structure is a pad diode, where the P-N junction is formed by bonding a P-type wafer to an N-type wafer. Figure~\ref{fig:structure_section}~(left) shows the cross-section of such a test structure. Ion beam implantation on the outer surfaces of the wafers form ohmic contacts between the silicon bulk and the metal used for contacts~\cite{yu_electron_1970}. The contact doping concentration at the wafer surface is targeted to be in the order of $N_C \approx 10^{20} ~ \text{cm}^{-3}$. The metal contacts are formed by a layered stack of chromium (as an adhesion layer), aluminium and gold (to prevent oxidation). On the N side guard rings are formed around the main central contact, in order to be able to reduce the leakage current across the diode edges. Experimentally we found that this was not necessary.
Diodes with contact edge lengths of 2mm, 4mm, 6mm, 8mm and 12mm are fabricated on the same wafer. A subset of the diodes on each wafer have circular metal openings centred on the contact, which will be used for TCT measurements. Figure~\ref{fig:structure_section}~(right) shows an image of a fully processed wafer pair where the arrangement of different diode sizes can be seen. As substrates we used 100mm (4") high resistivity N and P-type float zone wafers, as are used for fabrication of particle pixel detectors. The specifications of the used wafers are presented in table~\ref{tab:wafer_spec}.
\begin{table}
  \caption{Specifications of the wafers used for this production run. All are as specified by the wafer supplier, except the thickness which represents the value measured after polishing via CMP and the doping concentration which was calculated using~\cite{noauthor_pv_2022} based on the specified resistivity.}
  \label{tab:wafer_spec}

  \vspace{0.3cm}

  \resizebox{\textwidth}{!}{
    \begin{tabular}{l|rrrrr}
      \textbf{Type} & \textbf{Orientation} & \textbf{Size} & \textbf{Thickness} & \textbf{Resistivity} & \textbf{Doping} \\ \hline
      P (Boron) & <100> & $100 ~ \text{mm}$ & $500 \pm 7 ~ \mu\text{m}$ & $> 10 ~ \text{k}\Omega\text{cm}$ & $< 2\cdot10^{12} ~ \text{cm}^{-3}$ \\
      N (Phosphorous)& <100> & $100 ~ \text{mm}$ & $490 \pm 8 ~ \mu\text{m}$ & $7 - 10 ~ \text{k}\Omega\text{cm}$ & $(4 - 6) \cdot10^{11} ~ \text{cm}^{-3}$ \\
    \end{tabular}
  }
\end{table}
An overview of the processing steps is shown in table~\ref{tab:fabrication_plan}.
Processing was carried out in the Binnig and Rohrer Nanotechnology Center (BRNC) cleanroom of ETHZ/IBM as well as in the Center of MicroNanoTechnology (CMI) at EPFL.
Certain steps were carried out with external companies, namely the ion beam implantation with \emph{Ion Beam Services S.A.}~\cite{noauthor_ion_2016}, chemical-mechanical polishing (CMP) with \emph{Optim Wafer Services}~\cite{noauthor_optim_2022}, wafer-wafer bonding with \emph{EV Group GmbH}~\cite{ev_group_combond_2022} and laser dicing with \emph{Synova S.A.}~\cite{synova_sa_water_2022}.
\begin{table}
  \caption{Processing steps for the fabrication of the bonded test structures. Up to step 6 wafers were processed individually. Starting from step 8 wafers were processed as bonded pairs. Entries in \emph{italic} were carried out by external partners.}
  \label{tab:fabrication_plan}

  \renewcommand{\cellalign}{tl}
  \vspace{0.3cm}

  \centering
  \begin{tabular}{c|l|l}
    \textbf{Step} & \textbf{N-Wafer} & \textbf{P-Wafer} \\ \hline
    0 & Alignment marker etching & Alignment marker etching \\
    1 & 20nm PECVD cap oxide & 20nm PECVD cap oxide \\
    2 & Implant photolithography & Implant photolithography \\
    \emph{3} & \emph{\makecell{Ion beam implantation\\($8\cdot10^{15} \text{ cm}^{-2}$ @ 30keV)}} & \emph{\makecell{Ion beam implantation\\($6\cdot10^{15} \text{ cm}^{-2}$ @ 80keV)}} \\
    4 & Cap oxide removal & Cap oxide removal \\
    5 & \makecell{Doping activation\\($50 \text{ min}$ @ $1050^\circ$C)} & \makecell{Doping activation\\($70 \text{ min}$ @ $1050^\circ$C)} \\
    \emph{6} & \emph{Backside polishing with CMP} & \emph{Backside polishing with CMP} \\
    \emph{7} & \multicolumn{2}{c}{\emph{Wafer-wafer bonding}} \\
    8 & \multicolumn{2}{c}{N side metal evaporation} \\
    9 & \multicolumn{2}{c}{N side metal lift-off} \\
    10 & \multicolumn{2}{c}{P side metal evaporation} \\
    11 & \multicolumn{2}{c}{P side metal lift-off} \\
    \emph{12} & \multicolumn{2}{c}{\emph{Laser dicing}} \\
  \end{tabular}
\end{table}
The doping activation step at $1050^\circ \text{C}$ CMP of the backside ensures that the RMS roughness of the wafer surfaces to be bonded is $< 1 ~\text{nm}$~\cite{predan_direct_2020}. Additionally during the same process step, ca. $25 ~ \mu\text{m}$ of silicon was removed from the backside, in order to remove any contamination from the previous processing steps.

\subsection{Fabrication quality control}
\begin{figure}
  \centering
  \includegraphics[height=7cm]{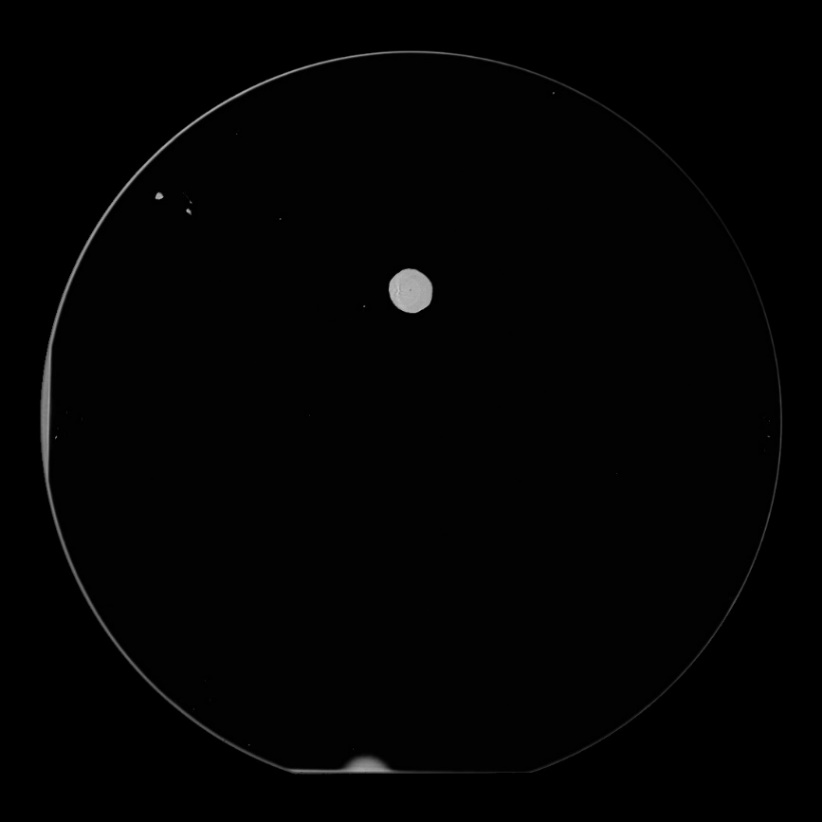} ~~
  \includegraphics[height=7cm]{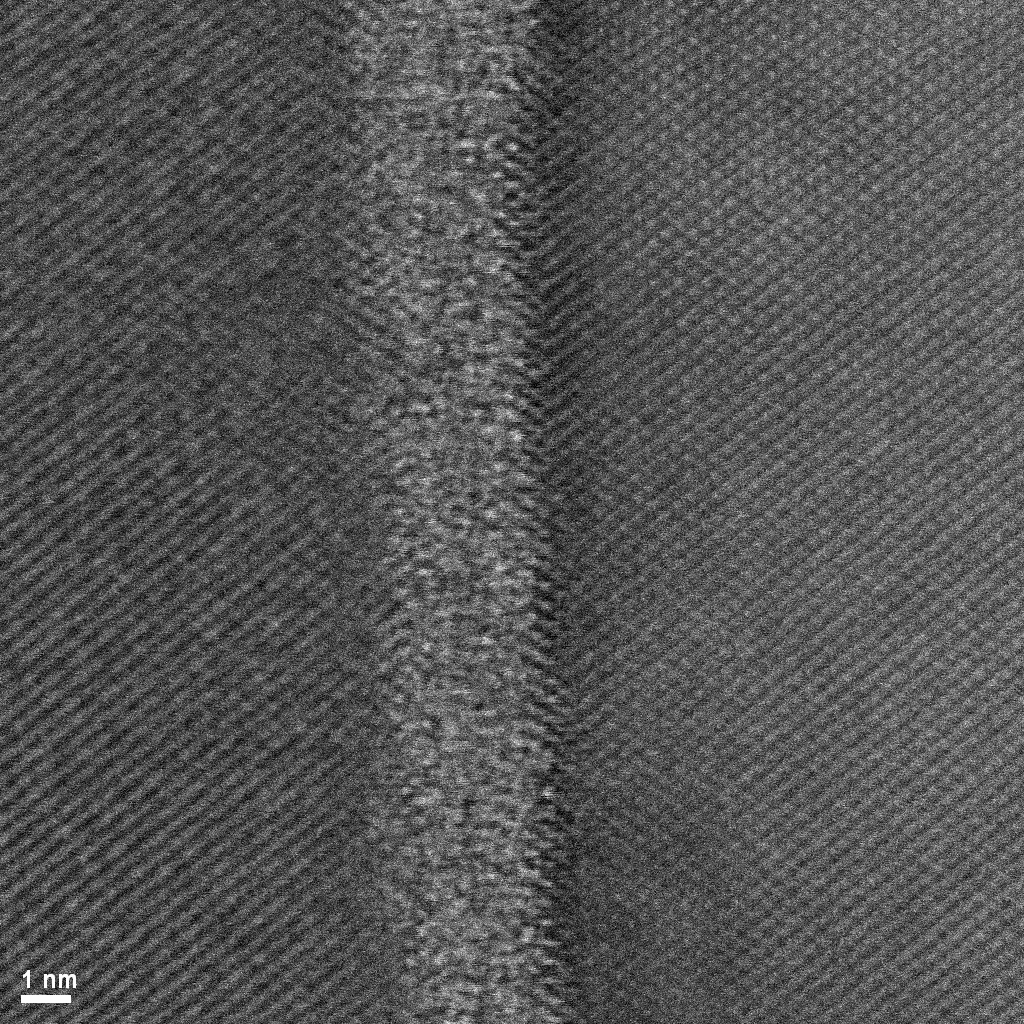}
  \caption{Left: SAM measurement of a bonded wafer pair indicating a good bond quality with one void due to a contaminating particle. ---- Right: STEM image of the bonding interface for the same pair, showing an amorphous layer in between the two crystalline silicon wafers.}
  \label{fig:bond_analysis}
\end{figure}
Quality control of the bonded pairs was carried out with scanning acoustic microscopy (SAM) measurements by the bonding company. The SAM measurements show a good bond quality with some voids due to contaminating particles. Figure~\ref{fig:bond_analysis}~(left) shows an example SAM measurement. In order to evaluate the width of the amorphous layer at the bond interface, scanning transmission electron microscopy (STEM) measurements were carried out at ScopeM -- ETH Zürich using a Hitachi HD 2700 STEM. Figure~\ref{fig:bond_analysis}~(right) shows the result of these STEM measurements.

\begin{figure}
  \centering
  \includegraphics[width=0.6\textwidth, clip, trim=0 0 0 1cm, valign=t]{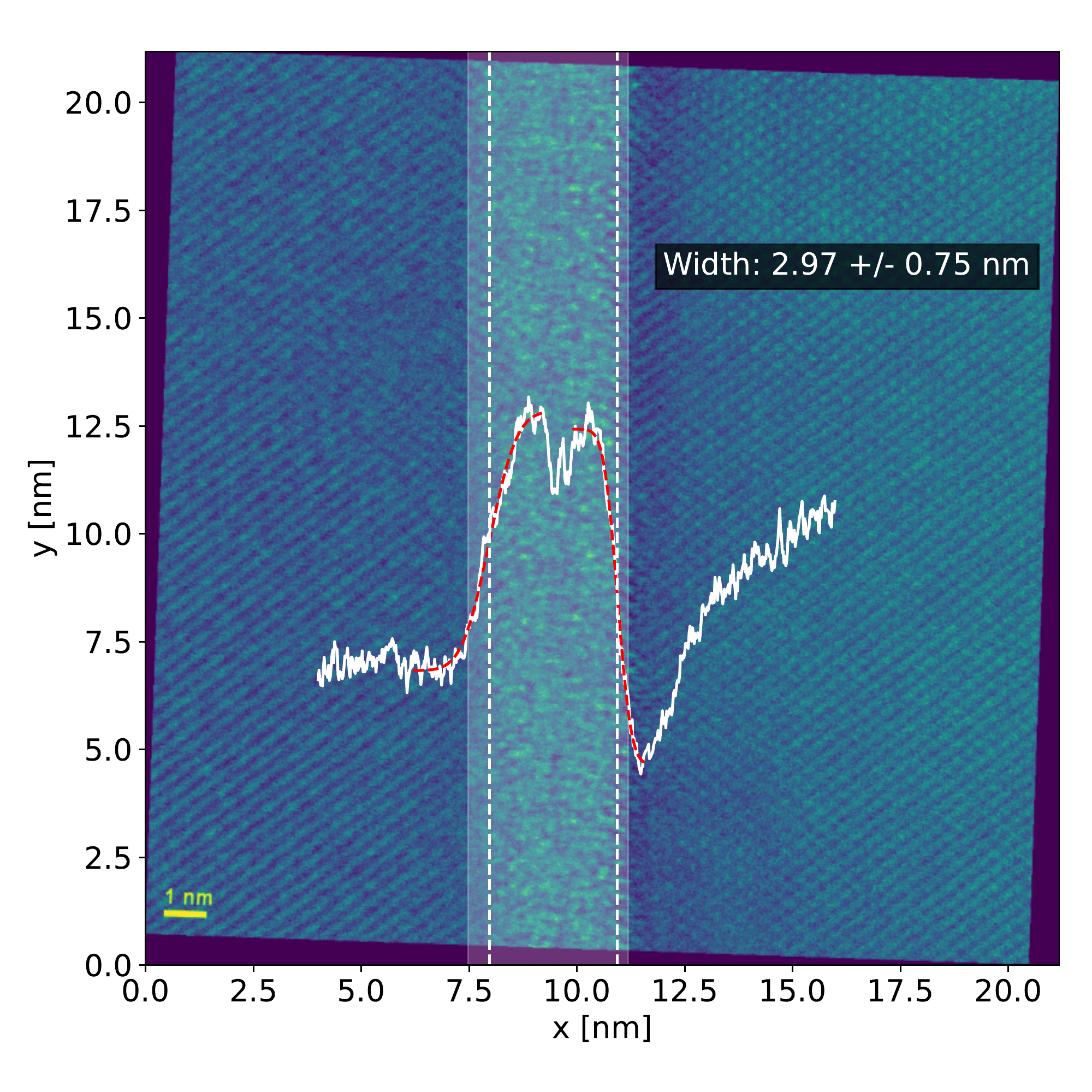}
  \caption{Estimation of the width of the amorphous layer. The overlayed white graph shows the average image intensity along the x axis. The two red dashed lines show the fit of the modified ERF for each edge of the amorphous layer. The estimated width of the amorphous layer is calculated based on the $\mu$ and $\sigma$ of the two fitted ERF.}
  \label{fig:interface_width}
\end{figure}

We estimated the width of the amorphous layer via numerical image processing. After rotating the image in order to have the interface vertically aligned, we calculated the average image intensity in function of the (x) axis perpendicular to the bonding interface. Figure~\ref{fig:interface_width} shows the resulting curve (in arbitrary units) overlaid onto the STEM image. The amorphous layer on average has a higher image intensity. The increase in intensity at the left (resp. right) side transition from crystalline to amorphous material is then fitted using a modified error function (ERF)
  $$I_{L,R}(x) = A\cdot\erf\left(\frac{x - \mu_{L,R}}{\sqrt{2}\sigma_{L,R}}\right) + I_{offset}.$$
We find the following width for the amorphous interface
  $$d_{amorph} = (\mu_{R} - \mu_{L}) \pm (\sigma_{R} + \sigma_{L}) = 2.97 \pm 0.75 \text{ nm}$$

The mask set of our process includes transfer length method (TLM) structures which we use for assessing the quality of the fabricated ohmic contacts~\cite{oussalah_comparative_2005}.
The contact resistivities measured with the TLM structures are in the order of $\rho_{Contact} \approx 0.05 ~ \Omega\text{cm}^2$.
Accordingly we expect contact resistance of the order of $1 \Omega$ for our smallest (2x2 mm) pad diodes.
This is negligible in context of the high resistivity substrates and confirms the fabrication of good ohmic contacts.

\section{Transient current technique measurements}

\begin{figure}
  \centering
  \includegraphics[width=0.43\textwidth, valign=c]{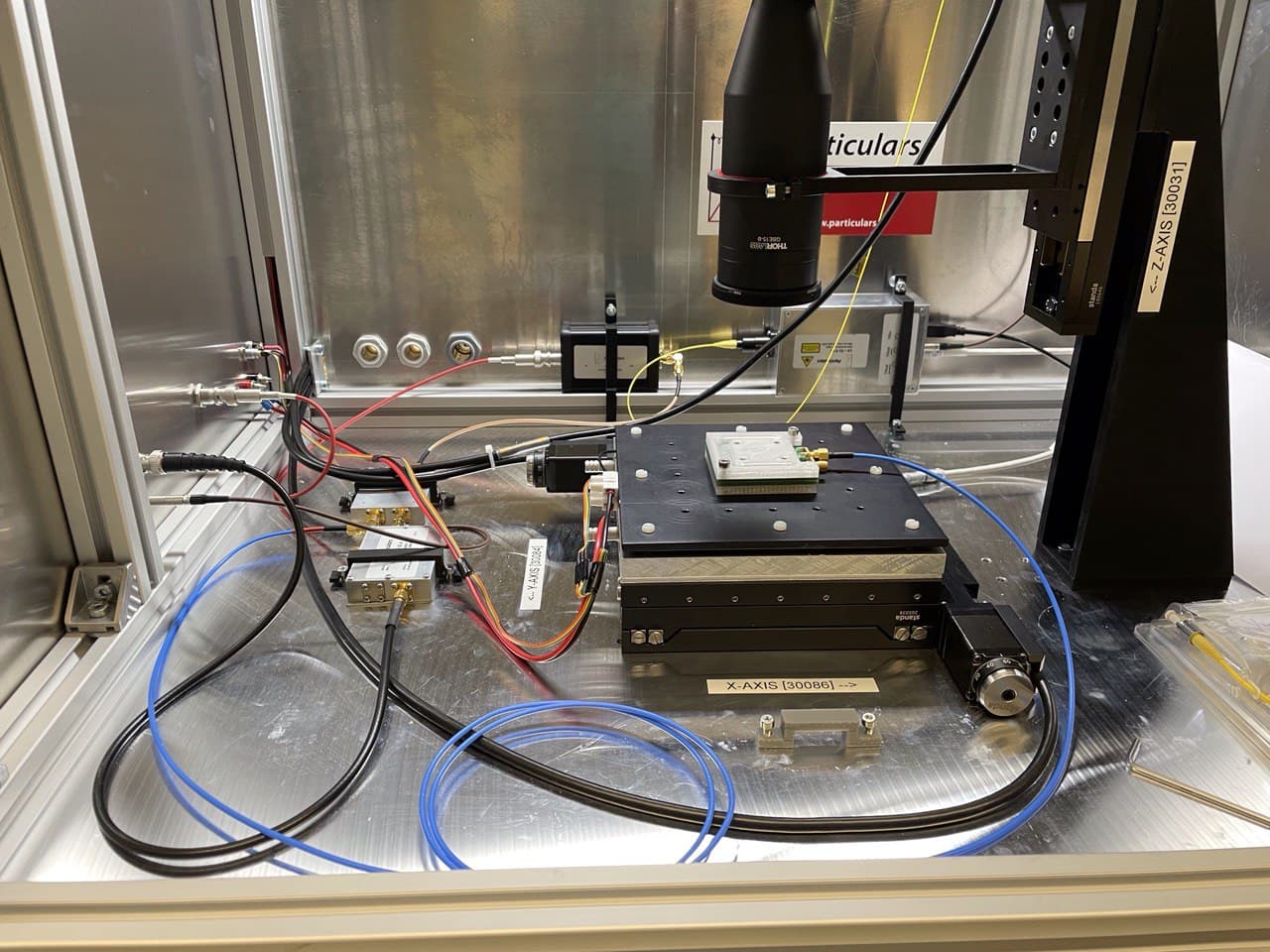} ~
  \includegraphics[width=0.55\textwidth, valign=c]{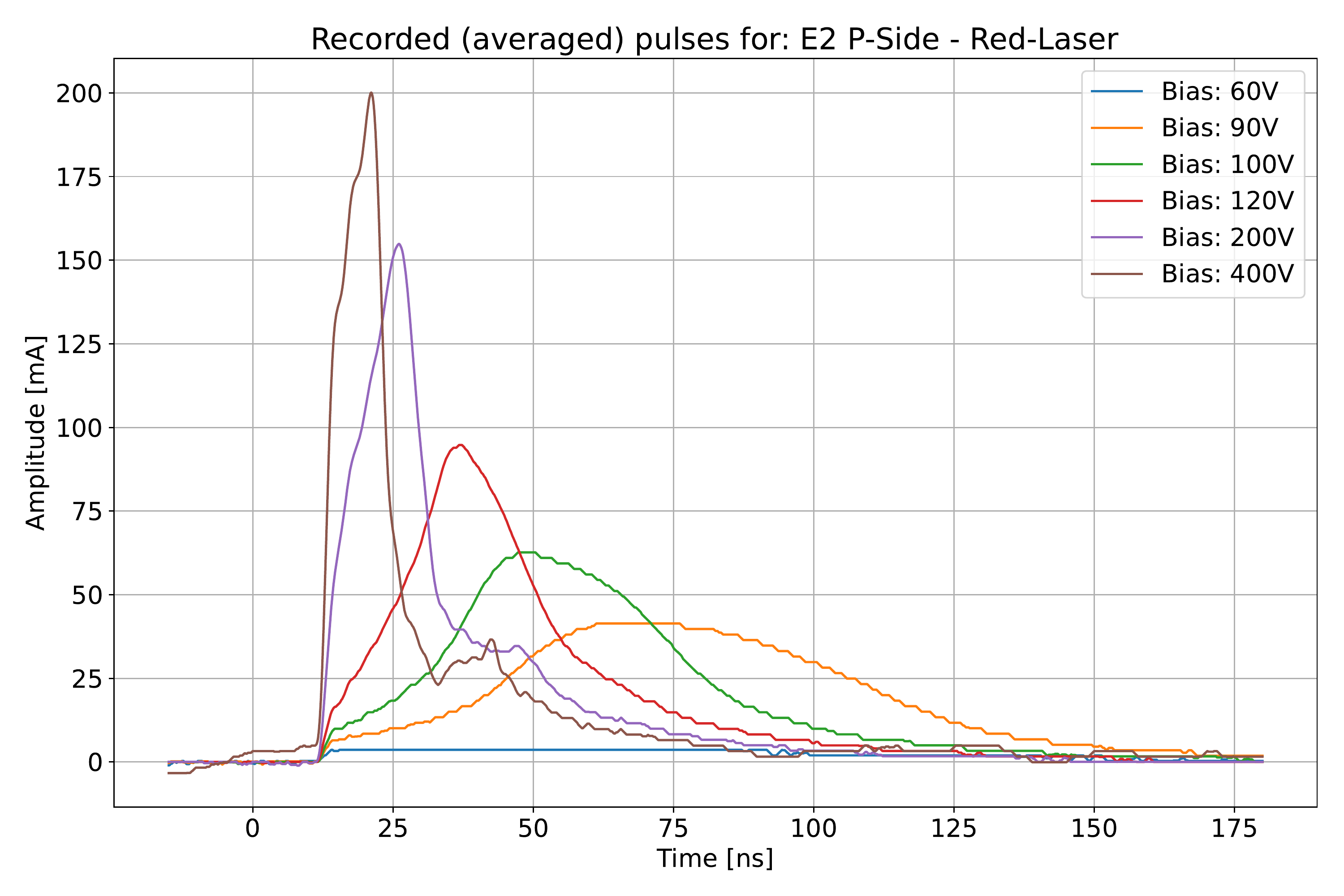}
  \caption{Left: TCT Setup with a test structure installed in a custom diode holder. --- Right: Example curves recorded when illuminating our samples with the red laser from the P side.}
  \label{fig:tct_example}
\end{figure}

Scanning TCT measurements were carried out using the \emph{Large Scanning TCT} setup acquired from Particulars~\cite{particulars_particulars_2021} shown in figure~\ref{fig:tct_example}~(left). Reverse bias is supplied by a Keithley 2410 SMU via the bias tee supplied by Particulars.
For our analysis we use two different lasers, one with a wavelength of 660 nm (red laser) and one with a wavelength of 1064 nm (IR laser).
Illumination with the red laser emulates the interaction of an alpha particle, in the sense that all the red light (the energy) is absorbed close to the sensor surface.
Illumination with the IR laser gives a minimum ionizing particle (MIP) like interaction, as the IR laser beam penetrates the entire sensor and is not fully absorbed. But while the MIP interaction is constant over the entire thickness of the sensor, the IR laser beam is attenuated and the laser intensity decreases exponentially within the sensor.

A pad diode with a contact size of $4 \text{ mm}$ was used for the in-depth analysis.
The diode has a 1mm diameter metal opening on each side and, accordingly, we carried out laser illumination from the P and the N side of the diode. Focusing of the TCT laser was carried out with a standard knife-edge scan making use of the edge of the metal opening~\cite{suzaki_measurement_1975}\cite{cindro_advanced_2015}.
Signals were recorded with Lecroy Waverunner~8104 oscilloscope with a bandwidth of 1~GHz and a sampling rate of 20~Gsps. For each recorded data point we averaged 1000 acquisitions in order to remove any (non correlated) noise.
We acquired reverse bias scans for bias voltages up to $|V_{Bias}| = 650 \text{ V}$ for all combinations of red and IR laser, and P side and N side illumination.
Figure~\ref{fig:tct_example}~(right) shows example curves recorded when illuminating the P side with the red laser.

In addition to the averaging of the curves using the oscilloscope, the recorded curves pass through two additional pre-process steps before analysis. To remove any constant offset in the signal, the mean signal value up to the trigger point is calculated (average over ca. $30 ~ \text{ns}$ of signal) and subtracted from the entire curve. The raw recorded signals have a superposed sinusoidal interference at a frequency of $f_{sin} \approx 290 ~ \text{MHz}$. We apply a notch filter with a centre frequency of $f_c = f_{sin}$ and a quality factor of $Q = 1$ to filter out this sinusoidal interference.

\subsection{Results} \label{sec:results}

\begin{figure}
  \centering
  \includegraphics[width=0.7\textwidth]{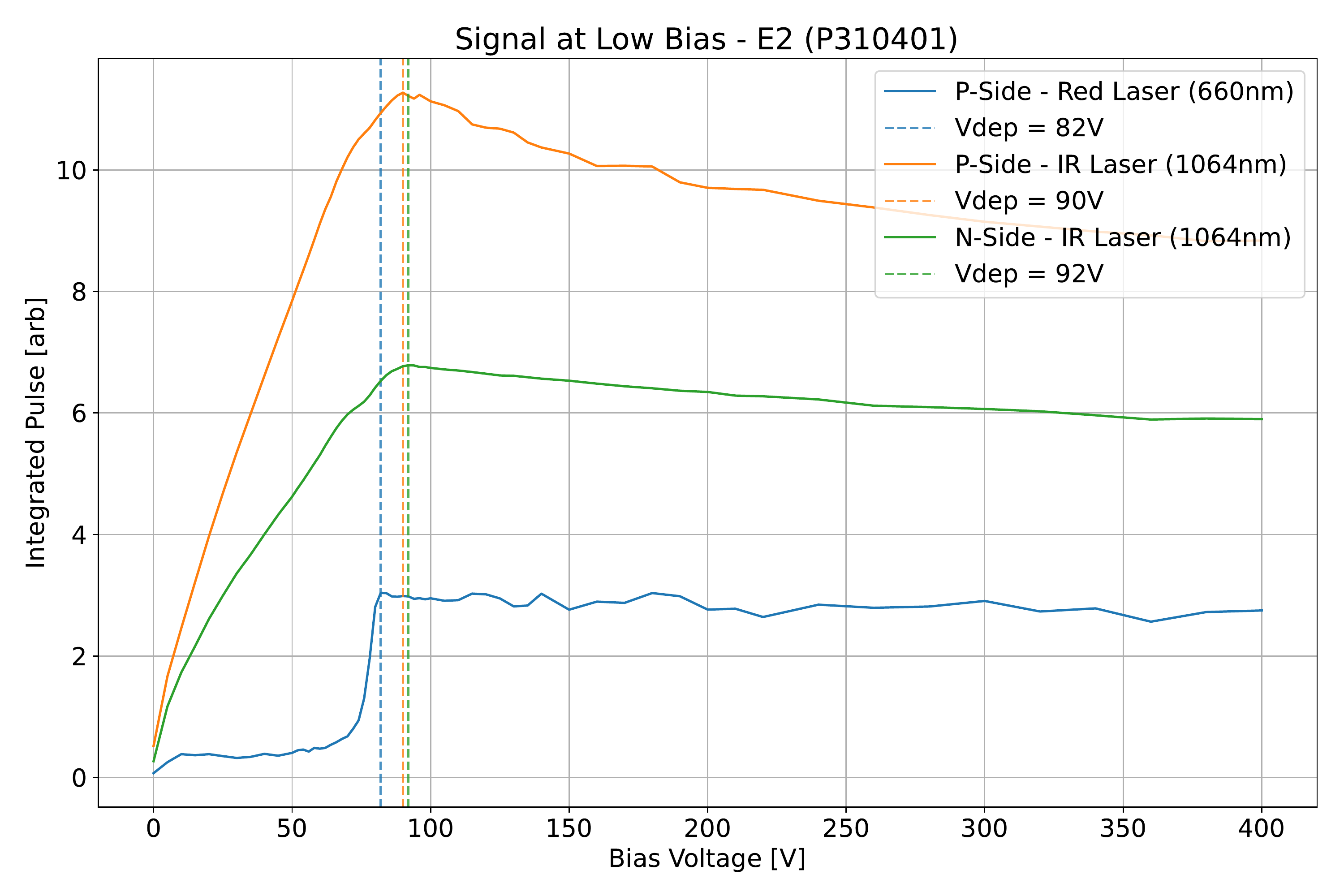}
  \caption{Integrated pulse amplitude in function of the applied bias voltage for different cases of TCT illumination. The decrease in collected signal beyond $100 \text{ V}$ in the case of IR laser pulses is currently unexplained and subject to further research.}
  \label{fig:tct_pulse_integral}
\end{figure}

TCT signals could be recorded in all four cases mentioned above, except in the case of illuminating the N side with the red laser.
Up to a bias voltage of $|V_{Bias}| = 650 \text{ V}$ no signal was observed in this case.
For the other three cases, the integrated pulse amplitude (i.e. the collected signal) as a function of the bias voltage is shown in figure~\ref{fig:tct_pulse_integral}.
It is immediately visible that in all three cases a maximum of collected charge is reached at bias voltages below $100 ~ \text{V}$.
In the case of the illumination with the red laser, this maximum is reached at ca. $V_{Red} \approx 82 ~ \text{V}$.
In the case of illumination with the IR, the maximum is reached at ca. $V_{IR} \approx 92 ~ \text{V}$.
The depletion region under reverse bias always grows from the P-N junction outwards. In the case of red illumination on the P side, we observe a plateau in signal being collected at $V_{Bias} > V_{Red}$. At $V_{Bias} < V_{Red}$ we see an exponential increase in collected signal.
This indicates that $V_{Red}$ is the voltage at which the electric field effectively reaches the surface of the P side, and therefore is equal to the voltage necessary to fully deplete the P wafer.
At this voltage, the N side does not appear to be fully depleted, as otherwise a signal would also be observed under red laser illumination from the N side.
One would expect the depletion region to grow further into the N side at $V_{Bias} > V_{Red}$ and thus the signal collected should also further increase beyond $V_{Bias} > V_{Red}$, due to the longer drift distance (Ramo's theorem~\cite{ramo_currents_1939}).
In conclusion, from the perspective of the charges collected on the P side, the drift distance seems to be limited to the width of the P wafer.

Under IR illumination the observed behaviour is more puzzling. Not only is the voltage at which the maximum is observed $V_{IR}$ higher, but also there is a very pronounced decrease in detected signals at $V_{Bias} > V_{IR}$. At $V_{Bias} >> V_{IR}$ the observed signal under IR illumination eventually reaches a steady state value. Under IR light illumination free charges are generated on both sides of the bonding interface. Therefore, even under the assumption that the bonding interface inhibits the drift of charge carriers, any increase in the depletion region in the N wafer should still lead to an increase in signal collected.
The observation, that the two voltages $V_{IR}$ and $V_{Red}$ are very close together, gives a hint, that any depletion of the N-wafer at $V_{Bias} > V_{Red}$ is very limited in size. All of this implies that there is a strongly asymmetric depletion behaviour between the P and N side of our structures. We attribute this behaviour to the presence of the bonding interface.
A total of six samples have been tested and the results from all six samples are consistent. The different samples showed variations in the voltage necessary for full depletion, which we attribute to differences in the precise substrate doping of each sample.
Upcoming edge TCT measurements should give us a more detailed and precise view of the exact depletion behaviour of these structures.

\section{Conclusions}
We fabricated custom pad diode test structures based on low-temperature covalent wafer-wafer bonding of high resistivity N to P-type silicon wafers. Implanting a highly doped layer at the surface of each wafer allows for good ohmic contacts. The inspection of the wafer-wafer bond with SAM measurements shows a good bond quality over the entire wafers. STEM measurements of the bonding interface show a ca. $3 ~\text{nm}$ thick amorphous layer at the bonding interface, which originates from the surface activation via an argon beam during the bonding process. Transient signals could be observed in our samples when illuminated using a TCT setup with red and IR laser pulses. Bias dependent measurements with the red laser indicate that a full depletion of the P side is possible, but not of the N side. The same measurements are consistent with the drift of charges not crossing the bonding interface. Together with the results from the IR laser measurements we observed a highly asymmetric depletion of the P side versus the N side. Further measurements will give more details about the exact shape of the depletion of our samples.

\acknowledgments
The authors thank the Cleanroom Operations Team of the Binnig and Rohrer Nanotechnology Center (BRNC) for their help and support during the fabrication process.
We further thank Dr.~P.~Zeng, Dr.~F.~Gramm and Dr.~E.~Barthazy of ScopeM for their support and assistance with the STEM analysis of our samples.
For data processing and plotting we rely on the Python programming language using the Numpy~\cite{harris_array_2020}, SciPY~\cite{virtanen_scipy_2020} and Matplotlib~\cite{hunter_matplotlib_2007} packages.

\bibliographystyle{JHEP}
\bibliography{INFIERI2021_Proceeding_JWuethrich}

\end{document}